\documentclass[reprint,
 amsmath,amssymb,
 aps,
nofootinbib
]{revtex4-2}

\usepackage{graphicx}
\usepackage{dcolumn}
\usepackage{bm}
\usepackage{hyperref}

\DeclareRobustCommand\onedot{\futurelet\@let@token\@onedot}
\def\onedot{. }
\def\eg{\emph{e.g}\onedot}

\begin{document}

\preprint{APS/123-QED}

\title{Universal Diffusion in Coulomb Crystals}

\author{M. E. Caplan} \email{mecapl1@ilstu.edu}
\author{D. Yaacoub}%
 \email{dyaacou@ilstu.edu}
\affiliation{%
 Department of Physics, Illinois State University, Normal, Illinois 61761, USA
}%

\date{\today}

\begin{abstract}
Diffusion coefficients for crystallized Coulomb plasmas are essential microphysics input for modeling white dwarf cores and neutron star crusts but are poorly understood. 
In this work we present a model for diffusion in Coulomb crystals. We show that melting and diffusion follow the same universal scaling such that diffusion is independent of screening. Our simulations show, contrary to prevailing wisdom, that the formation of vacancies is not suppressed by the large pressure. Rather, vacancy formation and hole diffusion is the dominant mode of self diffusion in Coulomb crystals. 
\end{abstract}

\maketitle

\textit{Introduction \textendash} 
Strongly coupled Coulomb (Yukawa) plasmas are found under diverse conditions, spanning dusty plasmas, inertial-confinement experiments, white dwarf cores, and neutron star crusts \cite{thomas1994plasma,Caplan_2017,Saumon_2022}. Microscopic transport properties of these plasmas are essential for high fidelity computational modeling. However, owing to the extreme temperatures and densities simulations are often the only way to access high energy-density matter at astrophysical conditions.

At sufficiently high pressure these plasmas crystallize, forming a Coulomb crystal. While many properties of Coulomb crystals are well studied, diffusion in the solid phase remains very poorly understood. Diffusion coefficients are particularly important input for modeling the direct transport of matter, such as separation in crystallizing white dwarfs and neutron stars, and also the microphysical origin of elastic phenomena in magnetars including crust breaking and creep \cite{Saumon_2022,thompson2017global}. These diffusion coefficients therefore have observable consequences for the light curves of crystallizing white dwarfs, accreting neutron stars, and bursting magnetars.

Astrophysical Coulomb crystals are under high pressure and past authors have generally assumed that vacancies do not form \cite{Hughto_2011,Caplan_2020}. Rather, it was thought that diffusion proceeded by `exchanges' of neighboring atoms on a lattice. These diffusive `hops' have been observed occurring in closed loops that are thermally activated, as nuclei must overcome the potential Coulomb barrier from their neighbors in order to complete these exchanges \cite{Hughto_2011}. \textit{In this work, we will show that the spontaneous formation of vacancy-interstitial defects and the subsequent migration of a hole through the lattice is the dominant mode of diffusion active in Coulomb crystals.}

These assumptions about diffusion are not without consequence. For example, diffusion coefficients are often taken to be zero in stellar evolution simulations of crystallized white dwarf cores, assuming that the core cannot further evolve once crystallized \cite[\eg][]{bauer2023carbon}. However, chemical separation processes may continue and even solid diffusion coefficients orders of magnitude smaller than liquid diffusion coefficients could still result in macroscopic evolution of the star \cite{Mckinven_2016,baiko2023liquidphase}.

Molecular dynamics (MD) simulations to calculate diffusion coefficients are computationally expensive, requiring long simulation times and large volumes, and so little progress has been made until now. Previously, only a few diffusion coefficients have been computed at a few specific screening lengths relevant for astrophysics while mixtures are almost entirely unstudied \cite{Hughto_2011,Caplan_2020}. Despite the computational expense associated with such simulations, advances in GPU supercomputing now allow for the efficient calculation of pairwise interactions in plasmas, enabling us to revisit and thoroughly characterize diffusion in Coulomb crystals with large-scale N-body simulations evolved for many millions of timesteps.

In this Letter, we derive a scale free law for diffusion in Coulomb crystals. We show that this law agrees with MD simulations over orders of magnitude of coupling strength and screening. We show that there is a fundamental connection between melting and diffusion related to the mobility of ions and together they follow the same universal scaling, such that diffusion can be expressed independent of screening. We also show that diffusion in Coulomb crystals is a highly correlated many-body process consisting of diffusive cascades initiated by hole formation and is well described by an Eyring model for the thermal activation of hole formation.

\begin{figure*}[t]
\begin{flushleft}
\includegraphics[width=0.24\textwidth, trim=40 35 40 5, clip=true]{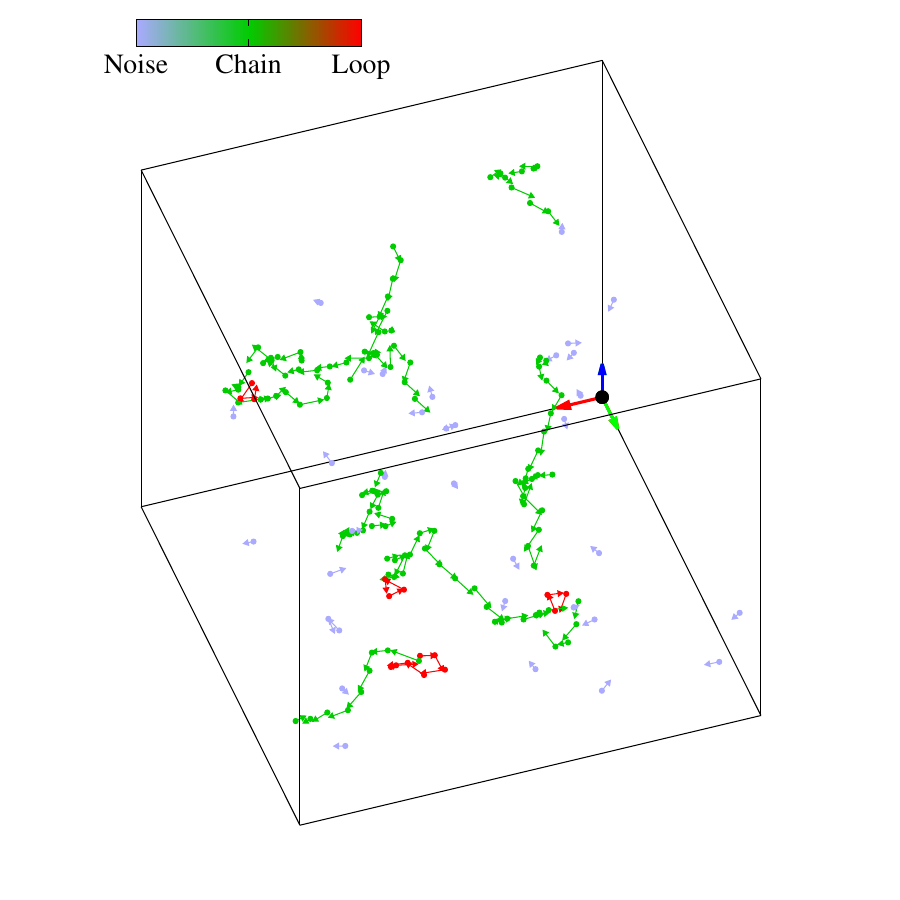}%
\includegraphics[height=3.45cm, trim=620 0 10 0, clip=true]{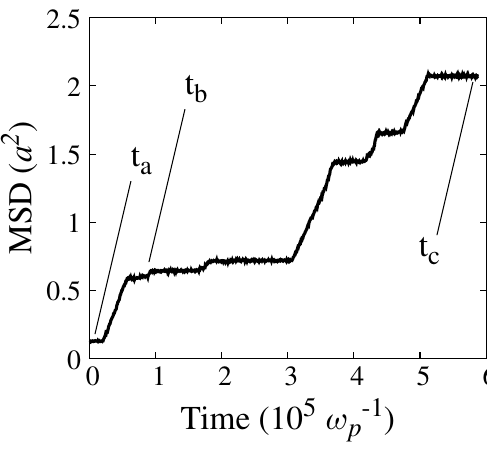}
\includegraphics[height=3.45cm, trim=5 0 250 0, clip=true]{fig1_MSD2.pdf}
\caption{\label{fig:MSD} (Left) Nuclei that have diffused a distance greater than $a_i$ in a short time have their displacement vectors shown. A diffusive cascade (green) is caused by the migration of a hole and interstitial following the formation of a vacancy-interstitial pair. Four closed loops (red) are visible near the chain, suggesting the motion of the chain also lowers the threshold for loop diffusion by adjacent nuclei. (Center) The MSD does not steadily grow as in a liquid, rather the MSD grows while there is an active vacancy-interstitial pair, but is otherwise quiescent and the slope (instantaneous $D^*$) is zero. (Right) Histograms of particle displacements make clear that nearest-neighbor exchanges occur while the instantaneous $D^*$ is nonzero, and that we are able to resolve many thousands of diffusive hops.}
\end{flushleft}
\end{figure*}

\textit{Coulomb Plasmas \textendash} The screened Coulomb potential models the two-body pairwise interaction of charged point particles in a plasma,

\begin{equation}
    V(r_{ij}) = \frac{e^2 Z_i Z_j}{r} e^{-r_{ij} /\lambda},
\end{equation}

\noindent with inter-particle separation $r_{ij}$ between two charges $e Z_i$ and $e Z_j$, taken to be nuclei in white dwarfs and neutron stars. Electrons are not explicitly included and instead provide screening $\lambda$, typically within a factor of a few of the inter-particle separation. The one-component plasma (OCP) consists of a single species of charge $Z$ and mass $m$ and is among the most well studied physical systems.  

The OCP is fully characterized by two dimensionless parameters, the coupling strength $\Gamma=e^2 Z^2 / a_i T$ and the screening length $\kappa = n_i^{-1/3} / \lambda$ with temperature $T$ and ion sphere radius $a_i = (4 \pi n_i /3)^{-1/3}$ using ion number density $n_i$. 
The unscreened OCP crystallizes at $\Gamma_{\rm crit} = 175.7$ \cite{baiko2022ab}. 
The melt line depends on $\kappa$, as the softening of the potential at high screening pushes crystallization to much lower temperature (higher $\Gamma$). 

A simple analytic expression for the melt curve that shows good agreement with numerical data up to $\kappa \approx 10$ is readily available in the literature \cite{vaulina2000scaling,PhysRevE.66.016404,Silvestri_2019}. The melting condition $\Gamma_{\rm M} (\kappa)$ can be found from the Lindemann melting criterion when using the frequency of the dust-lattice waves \cite{melandso1996lattice,vaulina2000scaling}

\begin{equation}\label{eq:melt}
\Gamma_{\rm M}(\kappa) \approx \Gamma_\mathrm{crit} \frac{ e^{\kappa} }{ (1 + \kappa + \frac{1}{2} \kappa^2)}    
\end{equation}

\noindent with $\Gamma_\mathrm{crit} = 175.7$ \cite{baiko2022ab}.
Fig. 2 in Ref. \cite{PhysRevE.66.016404} shows the success of Eq. \ref{eq:melt} in predicting the melt line; it is within a few percent for $\kappa \lesssim 1$ while overpredicting by at most ten percent when $\kappa \gtrsim 1$. With an expression for the melt line in hand, it is straightforward to define a `modified coupling parameter' $\Gamma^* = \Gamma / \Gamma_{\rm M}(\kappa)$ such that melting occurs at $\Gamma^* \approx 1$ for any screening length and the OCP is solid (liquid) above (below) this.

\begin{figure}[t]
\begin{flushleft}
\includegraphics[width=0.24\textwidth, trim=40 35 40 5, clip=true]{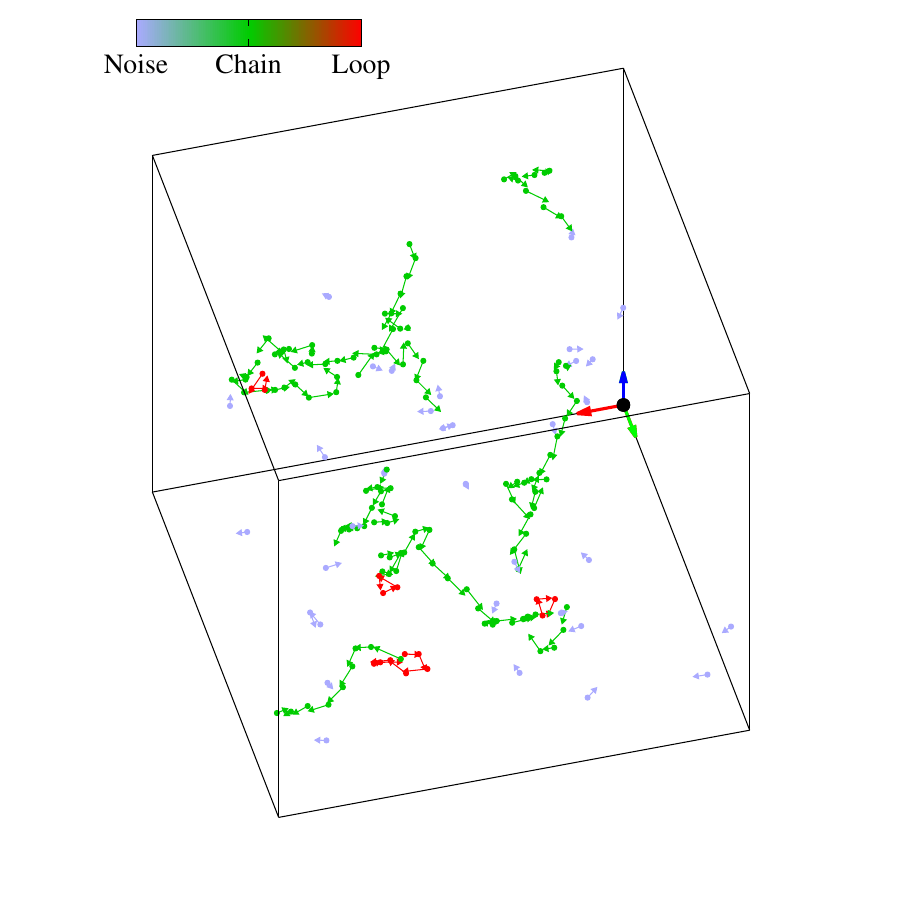}
\includegraphics[width=0.21\textwidth, trim=28 25 30 30, clip=true]{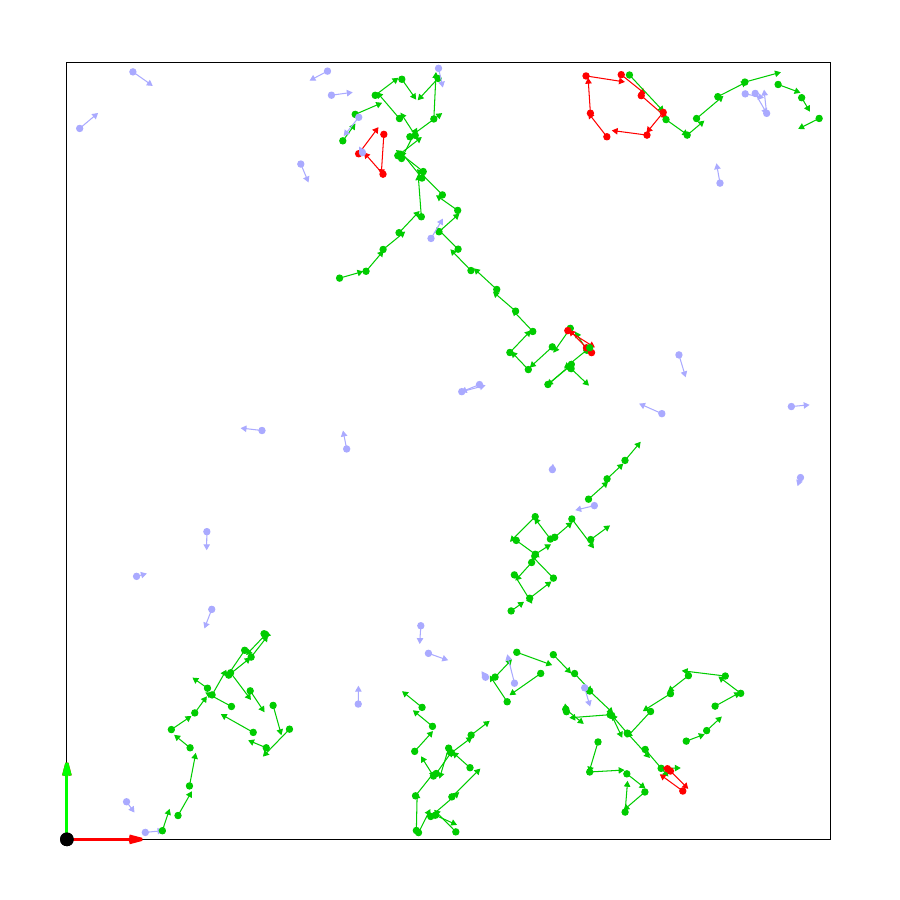}
\caption{\label{fig:chain2} (Left) Same as Fig. \ref{fig:MSD} but rotated for parallax. Rapidly switch between pg. 2 and 3 in a PDF viewer for effect. (Right) Projection on XY. Animations available in the SM.}
\end{flushleft}
\end{figure}

\textit{Diffusion Model \textendash } \textit{Our hypothesis in this work is that diffusion in the solid OCP can be fully characterized by this scale invariant $\Gamma^*$ that is relative to the melting temperature, \eg any OCP at twice $\Gamma_{\rm M}(\kappa)$ will have the same dimensionless diffusion coefficient regardless of $\kappa$.} Indeed, this has been shown for the OCP liquid \cite[e.g.][]{khrapak2018self,vaulina2000scaling}.
Diffusion in the solid phase similarly proceeds via thermally activated lattice site hops. However, these hops only occur when a neighboring lattice site is available, requiring one of two conditions: (1) the neighboring lattice site is vacant, or (2) the neighboring nucleus is simultaneously vacating its lattice site. This results in two modes of diffusion, the migration of holes resulting in diffusive cascades and closed loops of exchanges, shown on the left in Fig. \ref{fig:MSD} and Fig. \ref{fig:chain2}. Diffusion in solids is clearly a correlated many-body process, with no guarantee that techniques developed for liquids will hold. 

With a barrier of Coulomb energy $\Delta U$ we can express the diffusion coefficient in the probabilistic form, $D^* \propto \rm{exp}(- \Delta U/T) = \rm{exp}(-B \Gamma)$ \cite{Hughto_2011}. An appropriate Eyring model is therefore

\begin{equation}\label{eq:Dstar}
    D^* = \frac{A(\kappa)}{\Gamma} e^{-B(\kappa) \Gamma} = \frac{\alpha}{\Gamma^{*}} e^{-\beta \Gamma^{*} }
\end{equation}

\noindent by further defining $A \equiv \alpha \Gamma_{\rm M}(\kappa)$ and $B \equiv \beta / \Gamma_{\rm M}(\kappa)$ \cite{daligault2012diffusion}.

\textit{Molecular Dynamics \textendash} We compute diffusion coefficients from MD simulations of Coulomb crystals. Our formalism is the same as our past work \cite{Caplan_2021,Caplan_2022,caplan2023diffusion}. We use the IUMD v6.3.1 code to simulate $N=16000$ nuclei on a $20\times20\times20$ bcc lattice in a cubic box with periodic boundaries and periodic velocity rescaling to maintain the temperature and zero center of mass motion. In the large $N$ limit, the NVT and NVE ensembles are effectively equivalent for equilibrium MD; preliminary simulations comparing them showed agreement and this was verified in Ref. \cite{Hughto_2011} for similar systems.
Simulations are run with timestep $1/17 \, \omega_p$ (plasma frequency $~{\omega_p = (4 \pi e^2 Z^2 n_i / m )^{1/2}}$) for $10^7$ MD timesteps. This is an order of magnitude longer than the typical simulations in Ref. \cite{Hughto_2011}. Simulations span an order of magnitude in $\kappa$ up to about the triple point at $\kappa = 6.90$ because we are most interested in the bcc phase for astrophysics. We also omit the unscreened case due to the large required box sizes and the associated large computational cost.

We compute the normalized diffusion coefficient $D^* = D / \omega_p a^2 $ from a truncated mean square displacement (MSD) that omits ions that have not moved from their original lattice site via the Heaviside step function $\Theta$ \cite{Hughto_2011}

\begin{equation}
D(t)=\frac{\langle \Theta[|{\bf r}_j(t)-{\bf r}_j(0)|-R_c] |{\bf r}_j(t)-{\bf r}_j(0)|^2\rangle}{6 t}.
\label{D'}
\end{equation}

\noindent This is especially important at high $\Gamma^*$ where only a small number of nuclei undergo even one diffusive hop. In the limit ${t\to \infty}$, $D=D(t)$. We use a cut-off of $R_c = a_i$ \cite{Hughto_2011}. In Fig. \ref{fig:MSD} we show the MSD for $~{\kappa = 0.537}$ and $~{\Gamma = 185}$ (${\Gamma^* = 1.032}$). We show the distribution of particle displacements at three times (right). The distribution appears to evolve as it would in a liquid, with a roughly Poisson shaped envelope, except the nuclei are confined to lattice sites. However, (center) we see the MSD undergoes stochastic growth, with periods of quiescence separated by periods of roughly linear growth. These periods of quiescence are associated with the absence of any holes or defects in the simulation. At the start of a growth period, a vacancy-interstitial pair forms and rapidly separate. As these defects explore the volume they cause nuclei to move to adjacent lattice sites at a roughly constant rate, steadily increasing the MSD. When these defects encounter each other and annihilate diffusion is once again quenched. By inspection there is a hole present approximately 30 percent of the time at $\Gamma^*=1.032$, suggesting a mean hole density of $2\times10^{-5}\, n_i$. 

We visualize this process (left) by showing all nuclei that have been displaced a distance greater than $a_i$ in a small time interval as a displacement vector, pointing from their initial position to their final position. Complete animations are available in the supplemental materials (SM).\footnote{Preprint SM: \href{https://www.phy.ilstu.edu/~mecapl1/universal-diffusion/}{www.phy.ilstu.edu/$\sim$mecapl1/universal-diffusion/} } These are the nuclei that contribute positively to the calculation of the diffusion coefficient. Vectors are colored based on the distance to their nearest neighbors. The dominant mode of diffusion clearly involves long chains of nuclei, where a hole migrates through the lattice allowing nuclei to move onto a neighboring lattice site in a diffusive cascade. The underlying lattice planes are clearly visible from the symmetry of the path the hole takes. 
Meanwhile, nuclei that have exchanged lattice sites in a closed loop are shown in red. Four loops are observed in this time interval, one of eight nuclei and three of three nuclei. All are adjacent to the chains, suggesting that the passing of the chain may have catalyzed the loop. A plausible interpretation is that the presence of the vacancy lowers the activation energy for lattice site hops among adjacent nuclei, even if those nuclei do not move into the vacancy. Lastly, those nuclei that have no neighbors that have diffused greater than $a_i$ are shown in blue. These are thermal noise; these nuclei have shown maximum displacement on their own lattice site. They are included because many appear adjacent to the chain, and so large thermal oscillations by neighbors may also be key to briefly lowering the local activation energy to trigger a hop. In \href{https://www.phy.ilstu.edu/~mecapl1/universal-diffusion/SM2.mp4}{SM2}, we simulate at the same conditions as Fig. \ref{fig:MSD} but with a single nucleus removed to induce a hole; no quiescence is observed, through periods of enhanced diffusion are observed where an additional hole forms.

\begin{figure}[t]
\includegraphics[width=0.49\textwidth]{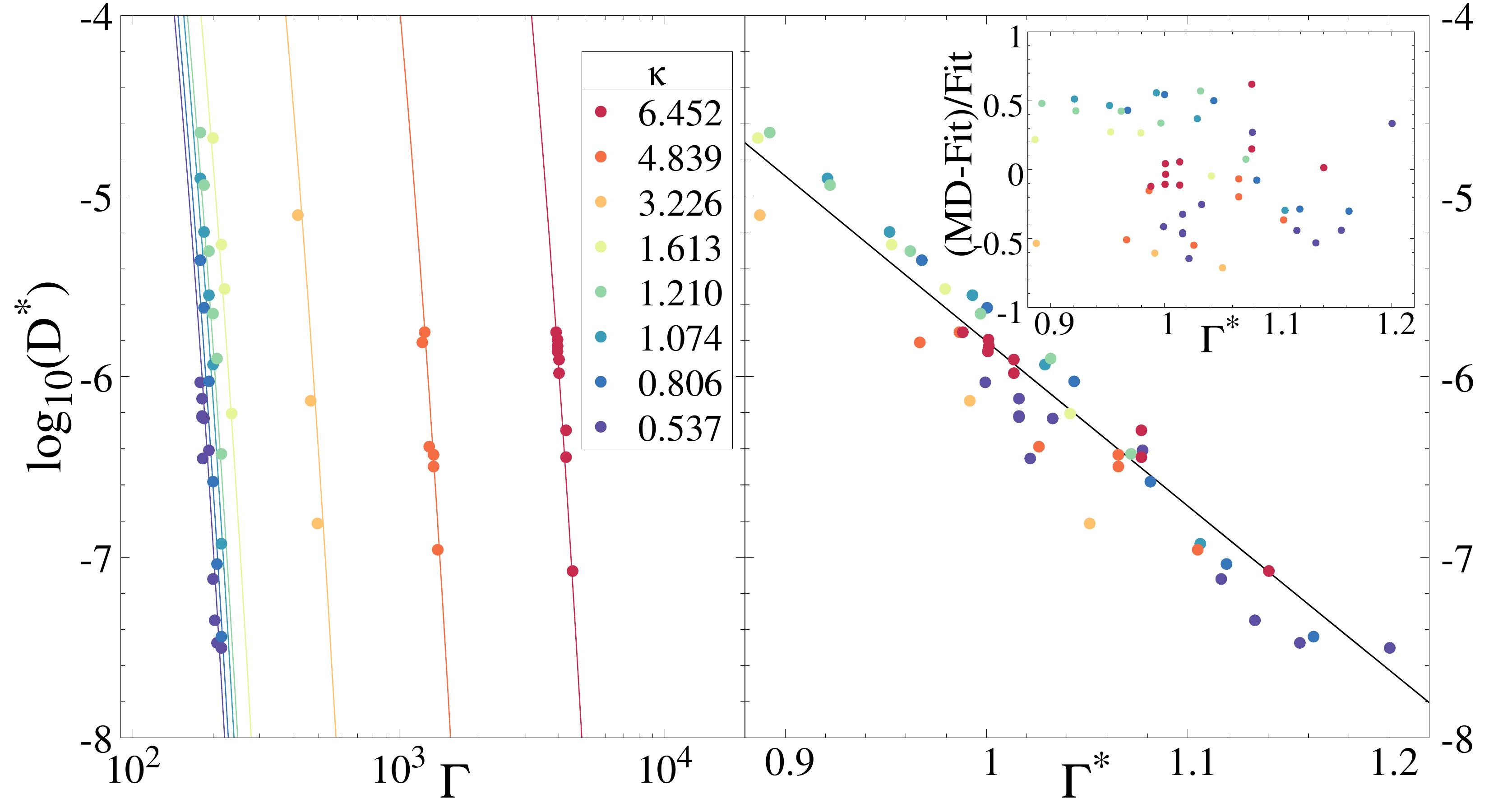}
\caption{\label{fig:Dstar} (Left) Diffusion coefficients for our MD and fits from Eq. \ref{eq:Dstar} collapse to a single curve (right) when $\Gamma$ is transformed to be relative to the $\Gamma_{\rm M}$ with the modified $\Gamma^*$.}
\end{figure}

In Fig. \ref{fig:Dstar} we show $D^*$ computed from MD. On the left we show the MD data and the best fit line from Eq. \ref{eq:Dstar}, $\alpha = 787$ and $\beta = 20.0$, fitting all our MD data simultaneously. This is similar to the Arrhenius fit with $B\approx 0.11$ given in Ref. \cite{Hughto_2011}. On the right we apply the transformation to the modified $\Gamma^*$ and the eight fit lines collapse to the central black line. 
The residual (inset) shows that typical errors are of order a few $10^{-1}$. Our model works best near melting and near $\kappa\sim1$, which is easy to interpret. To start, Eq. \ref{eq:melt} best predicts the melt line for small $\kappa$ but can have up to a $10^{-1}$ error itself. While $\Gamma^* \approx 1$ should be melting, some of our systems remain solid up to $\Gamma^* \approx 0.9$, though these systems could be superheated and only metastable.  Small uncertainty in the melting temperature is likely amplified by the exponential $B(\kappa)$ in our model; with $\beta \approx 20$, moving the melt from $\Gamma^*=1$ to $\Gamma^* = 1.05$ results in an extra e-folding. 

It is difficult to separate uncertainty from the melt line and MD stochasticity. Repeated simulations near $\Gamma^* =1$ show $10^{-1}$ variation in $D^*$ from MD stochasticity. MD stochasticity dominates the error for $\Gamma^* >1.1$ where we only observe one or two holes in these simulations.
At the highest $\Gamma^*$, some underprediction of $D^*$ from the MD may also be expected. At very high $\Gamma^*$ we do not expect a single diffusive hop during our MD runs, which is obviously a finite size effect. If we expect the infinite system to have some equilibrium hole density, then a counting bias is introduced when the equilibrium number of holes is fewer than one. When our simulations are sufficiently small that they contain at most one hole between periods of quiescence the hole-interstitial density will be significantly higher than average while the hole exists, making it easier for the hole to find its interstitial and annihilate so the lifetime of a hole is shorter than would be expected in a larger system.   

Future work should study the equilibrium hole density and compute activation energies for hole formation and hole migration. In \href{https://www.phy.ilstu.edu/~mecapl1/universal-diffusion/SM3.mp4}{SM3}, we show animations of the MSD and diffusion for $\Gamma^*=1$ and $\Gamma^*=1.07$ with $\kappa=0.537$ and $\kappa=6.452$. While these simulations show almost equal diffusion coefficients, at higher $\kappa$ there is clearly a higher hole density but slower hole diffusion, while at lower $\kappa$ the hole density is lower but hole diffusion is faster. This suggests that the activation energy for hole formation and hole migration may not scale the same, and the model presented in this work may best be thought of as a first order approximation.

\textit{Conclusion \textendash} In this work we have demonstrated that diffusion in Coulomb crystals has a universal character, so that regardless of screening length the temperature relative to the melting temperature is the only relevant parameter for determining diffusion coefficients. We have further shown that this is due to the formation of vacancy-interstitial defects, and that vacancy diffusion is actually the dominant mode of self-diffusion active in Coulomb crystals, even despite the astronomically high pressures and defect energies associated with them. 

It may be surprising that a rescaled Eyring model predicts $D^*$ as well as it does. In liquids, the activation energy in the Eyring model may simply be that for a `hop' between amorphous cages of neighbors. We have shown how diffusion in solids is a correlated many-body process and find evidence for multiple independent activation energies with different scaling. In \href{https://www.phy.ilstu.edu/~mecapl1/universal-diffusion/SM3.mp4}{SM3} one can observe that the equilibrium hole density is not invariant with $\Gamma^*$ but has a strong $\kappa$ dependence, and likewise the activation energy for hole (and interstitial) diffusion scales strongly with $\kappa$. Despite this non-universality, which should be explored in more detail, $D^*$ in our model still agrees within a few times $10^{-1}$.

Many outstanding problems remain. To begin, molecular dynamics simulations can only explore diffusion near the melting temperature, and simulations in this work were only able to explore down to about $0.8 \, T_{\rm{m}}$. Still, we have shown diffusion in solids near melting is only two orders of magnitude smaller than a liquid at melting \cite{caplan2023diffusion}. This is sufficient for many purposes, as diffusion coefficients near melting are most obviously important for separation processes occurring at solid-liquid interfaces such as the white dwarf core-mantle boundary and the accreting neutron star crust-ocean boundary. Beyond this point diffusion quenches and it becomes prohibitively expensive to simulate long enough to see even one diffusive event. Given that the formation of a hole is a many-body process, it is possible that at very large $\Gamma^{*}$ the diffusion may be more strongly suppressed and better modeled by a modified Eyring equation with $\Gamma^{*}$ replaced by $( \Gamma^{*} )^{n}$. 

This work motivates future work to determine the OCP melt line to high precision at high resolution in $\kappa$. The form of the melt line used in this work is empirical and incredibly useful, but is a major source of uncertainty due to the exponential dependence on $\Gamma_{\rm M}(\kappa)$ in Eq. \ref{eq:Dstar}. With precise knowledge of the melt line our diffusion model can be refined to high precision with relatively little investment in MD of diffusion itself. 

Direct applications of our model to astrophysics are immediate. This model could be implemented in a stellar evolution code such as MESA straightforwardly \cite{jermyn2023modules}, but we also emphasize that more work is still needed for accurate diffusion coefficients for realistic Coulomb crystals. For example, the abundance of defects and the characteristic crystal domain size almost certainly plays an important role in radiating and absorbing holes, and the sensitivity to grain boundary orientations must also be explored \cite{Caplan_2020,baiko2022ab}. Pycnonuclear fusion likewise provides a mechanism for readily generating holes, and is almost certainly relevant to the evolution of accreting neutron star crusts \cite{chamel2008physics}. For mixtures in white dwarf cores and accreting neutron star crusts, this also likely has a composition dependence \cite{Mckinven_2016,caplan2021cooling}. Furthermore, low $Z$ impurities may act similar holes and many phase diagrams suggest that even relatively pure crystals that precipitate may have impurity number fractions of $10^{-2}$ to $10^{-3}$ \cite{caplan2021cooling,baiko2022phase}. All these factors are likely highly dependent on the site, composition, and thermal history and should be explored in future work. Beyond astrophysics, the model here can be applied to low pressure systems such as dusty plasmas; if experiments could artificially induce single holes then the defect mobility should be straightforward to measure and compare.

\begin{acknowledgments}
The authors thank Yuri Levin and Ashley Bransgrove for discussion. 
Financial support for this publication comes from Cottrell Scholar Award \#CS-CSA-2023-139 sponsored by Research Corporation for Science Advancement. This work was supported by a grant from the Simons Foundation (MP-SCMPS-00001470) to MC. This research was supported in part by the National Science Foundation under Grant No. NSF PHY-1748958. 
M.C. thanks the KITP for hospitality and acknowledges support as a KITP Scholar. This research was supported in part by Lilly Endowment, Inc., through its support for the Indiana University Pervasive Technology Institute.

\end{acknowledgments}




\nocite{*}


%

\end{document}